\documentclass[12pt,reqno]{amsart}
\usepackage[utf8]{inputenc}
\usepackage[T2A]{fontenc}      
\usepackage{amssymb}
\usepackage{latexsym}
\usepackage{amsfonts}
\usepackage{amsmath}
\usepackage{amsthm}
\usepackage{mathrsfs}
\usepackage{amssymb}
\usepackage{stmaryrd}
\usepackage{graphicx}
\usepackage{epstopdf}
\usepackage{textcomp}
\usepackage[margin=1in]{geometry}
\usepackage{booktabs}
\usepackage{longtable}
\usepackage{mathtools}
\usepackage{makecell}
\usepackage{algorithm}
\usepackage{algpseudocode}
\floatname{algorithm}{}

\renewcommand{\algorithmicrequire}{\textbf{Input:}}

\usepackage{hyperref}
\makeatletter
\@namedef{subjclassname@2020}{\textup{2020} Mathematics Subject Classification}
\makeatother

\newtheorem{theorem}{Theorem}

\theoremstyle{definition}

\newtheorem{assumption}{Assumption}
\newtheorem{remark}{Remark}

\begin{document}
\title[Relative utility bounds for empirically optimal portfolios]{Relative utility bounds for empirically optimal portfolios}
\author{Dmitry B. Rokhlin}
\address{I.I.\,Vorovich Institute of Mathematics, Mechanics and Computer Sciences  and Regional Scientific and Educational Mathematical Center of Southern Federal University}
\email{dbrohlin@sfedu.ru}
\thanks{The research is supported by the Russian Science Foundation, project 17-19-01038}
\begin{abstract}
We consider a single-period portfolio selection problem for an investor, maximizing the expected ratio of the portfolio utility and the utility of a best asset taken in hindsight. The decision rules are based on the history of stock returns with unknown distribution. Assuming that the utility function is Lipschitz or H\"{o}lder continuous (the concavity is not required), we obtain high probability utility bounds under the sole assumption that the returns are independent and identically distributed. These bounds depend only on the utility function, the number of assets and the number of observations. For concave utilities similar bounds are obtained for the portfolios produced by the exponentiated gradient method. Also we use statistical experiments to study risk and generalization properties of empirically optimal portfolios. Herein we consider a model with one risky asset and a dataset, containing the stock prices from NYSE. 
\end{abstract}
\keywords{Portfolio selection; Relative utility; Statistical learning; Empirical utility; Generalization bounds}
\subjclass[2020]{91G10, 68Q32}
\maketitle
\section{Introduction}
We consider a single-period portfolio selection problem, where the decision rules are based on the history of stock returns. It is assumed that the returns are independent and identically distributed, but their distribution is unknown. We represent investor's preferences by an expected utility and use  the sample average approximation (SAA) (see, e.g., \cite{Kim2015}) for the solution of the related expected utility maximization problem. In the terminology of the statistical learning theory our main goal is to obtain high-probability bounds (generalization bounds or utility bounds) for the difference between the optimal utility value and the true utility of the empirically optimal portfolio (estimation error), as well as for the difference between the true utility and the empirical utility for such portfolio. 

Let us mention two specific features of the problem under consideration, which make some difficulties in an application of standard results. First, some classical utility functions, like the power function, are neither bounded nor globally Lipschitz. Second, most classical models, like the Black-Scholes, assume that the returns are unbounded. Similar unbounded problems appear in general learning theory: see \cite{Cortes2019} and a lot of references therein. They require some additional assumptions, problem reformulations and the development of special tools.

In the present paper we pass to the \emph{relative} utility maximization, where the objective function equals to the expected ratio of the utility $u$ of some portfolio to the utility of the best portfolio for the returns, which are known in hindsight. This allows to avoid any assumption on the returns, besides the i.i.d. hypothesis. As for $u$, we assume that it belongs to the class of positive, non-decreasing functions, satisfying the global Lipschitz or H\"{o}lder condition, and some specific condition, regarding its behavior at zero and infinity. The power function satisfies these assumptions.  For the same problem with a concave utility function we study the estimation error for the portfolio produced by the stochastic version of the exponentiated gradient algorithm of \cite{Kivinen1997}. 

The obtained utility bounds contain only those quantities, which are known for the investor: the number of return observations; the number of stocks; constants, related to the utility function; and a data-dependent quantity in the case of the exponentiated gradient algorithm: Theorems \ref{th:1} -- \ref{th:3}.

Passing to the relative utility certainly affects investor's attitude towards risk. In the case of one risky asset it appears, that an investor with the relative utility is more risk averse than in the case of the ordinary utility. However, in the case of multiple risky assets our empirical results show that the situation can be the opposite. Furthermore, we present simple statistical experiments demonstrating that typically it is impossible to get a reliable estimate of the optimal portfolio on the base of daily historical observations. A related phenomenon, which was mainly demonstrated for the risk-return modeling of investor's preferences, is known as the fragility of SAA in  portfolio optimization: see \cite{Ban2018} and references therein.  

Let us mention some papers, considering single-period portfolio selection problems in the statistical learning framework. In \cite{DeMiguel2009,Gotoh2011}, the authors studied the influence of the portfolio constraints on the out-of-sample performance. The papers \cite{Gotoh2011,Gotoh2012} presented out-of-sample bounds for the loss probabilities of the portfolios, satisfying some empirical VaR- and CVaR-type constraints. The regularization and cross validation methods were applied to the mean-variance and mean-CVaR problems in \cite{Ban2018}. One can also find in \cite{Ban2018} several other references to the works, considering the regularization methods. In \cite{Bazier2020} the authors considered an expected utility maximization problem with side information and applied a regularization to obtain out-of-sample guarantees for the certainty equivalent of the out-of-sample portfolio value. 

The rest of the paper is organized as follows. In Section \ref{sec:2} we state the problem and mention the consistency of the  SAA method. Section \ref{sec:3} contains the main result of the paper: Theorem \ref{th:2}, which gives upper bounds for the expected maximum of an empirical process, associated to the relative utility function. The Lipschitz and H\"older cases are studied separately. In both cases we consider the Rademacher complexity of the class of relative utility functions, parametrized by the portfolio weights. In the Lipshitz case this quantity is estimated by the Talagrand contraction lemma and the Massart lemma, in the H\"older case we consider the packing numbers and the Dudley entropy integral. The obtained estimates directly lead to high-probability utility bounds via the concentration inequalities. Section \ref{sec:4} presents similar bounds for the portfolios produced by the stochastic exponentiated gradient algorithm of \cite{Kivinen1997}. Here we combine its online version with the online-to-batch conversion scheme: see \cite{Shalev2012}.

Sections \ref{sec:5} and \ref{sec:6} deal with  statistical experiments, related to the analysis of risk and generalization properties of empirically optimal portfolios. Section \ref{sec:5} considers the case of one risky asset, obeying the discrete Black-Scholes model, while in Section \ref{sec:6} we analyze a dataset, containing daily stock returns form NYSE. The conclusions are already briefly described above. Here we additionally indicate the utilized solution methods for the empirical utility maximization problems. In Section \ref{sec:5} the problem is one-dimensional, and it is solved simply via the bisection method. In Section \ref{sec:6} we propose a greedy modification of the stochastic exponentiated gradient algorithm to solve the correspondent is multidimensional problem. For logarithmic utility the results are compared with \cite{Borodin2000,Gyorfi2012}. The code for Sections \ref{sec:5}, \ref{sec:6} is available at \url{https://github.com/drokhlin/Relative_utility_bounds_code}.

\section{Problem formulation} \label{sec:2}
Let $(s_k^1,\dots, s_k^d)$ be strictly positive prices of $d$ assets (stocks) at time moments $k=0,\dots, n+1$, and let $r_k^j=s_i^j/s_{k-1}^j$, $j=1,\dots, d$, $k=1,\dots, n+1$ be the total daily returns (price relatives). At time $n$ an investor distributes his wealth $X_n=1$ between these assets based on the price history $(r_1,\dots,r_n)$. In other words, he selects a portfolio $(\gamma_n^1,\dots,\gamma_n^d)$, where $\gamma_n^j(r_1,\dots,r_n)\ge 0$ is the number of units of the asset $j$ to be bought. So, the wealth will be distributed between $d$ assets  in accordance with the fractions (or weights)
\[\nu_n=\left(\frac{\gamma_n^j s_n^j}{X_n}\right)_{j=1}^d\in\Delta=\left\{z\ge 0:\sum_{j=1}^d z_j=1\right\}. \]
At time $n+1$ the wealth becomes 
\[X_{n+1}=\langle\gamma_n,s_{n+1}\rangle=\langle\nu_n,r_{n+1}\rangle.\]
By $\langle a,b\rangle$ we denote the usual scalar product in $\mathbb R^d$.

Our standing assumptions concern the investor utility function and the returns.
\begin{assumption} \label{as:1} 
Investor's utility function $u:(0,\infty)\mapsto (0,\infty)$ is non-decreasing and continuous. 
\end{assumption}
\begin{assumption} \label{as:2} 
The return vectors $(r_k^1,\dots,r_k^d)$, $k=1,\dots,n+1$ are independent and identically distributed.
\end{assumption}

Consider the single-period optimization problem
\begin{equation} \label{1.1}
U(\nu)=\mathsf E f(\nu,r_{n+1}):=\mathsf E \frac{u(\langle\nu,r_{n+1}\rangle)}{u\left(r_{n+1}^* \right)}\to\max_{\nu\in\Delta},\qquad r_{n+1}^*:=\max_{1\le j\le d} r^j_{n+1}. 
\end{equation}
The objective function $U(\nu)$ of this problem equals to the expected ratio of the utility $u$ of some portfolio $\nu$ to the utility of the best portfolio taken in hindsight, that is, under the assumption that the values $r_{n+1}$ are known.
In the latter case the investor simply takes an asset with the largest return. Since $u$ is non-decreasing, the relative utility $f$ takes values in $(0,1]$. The set $\Delta$ is compact and the function $U$ is continuous, as follows from the continuity of $\nu\mapsto f(\nu,r)$ and the dominated convergence theorem. Hence an optimal solution $\nu^*$ of (\ref{1.1}) exists.

It is natural to consider the empirical utility maximization problem
\begin{equation} \label{1.2}
\widehat U_n(\nu)=\widehat f_n(\nu,r_{n+1})=\frac{1}{n}\sum_{k=1}^n \frac{u(\langle\nu,r_k\rangle)}{u(r^*_k)}\to\max_{\nu\in\Delta}.
\end{equation}
Clearly, this problem also has an optimal solution $\widehat \nu_n$.

Furthermore, consider the empirical process $\nu\mapsto G_n(\nu)=\widehat U_n(\nu)-U(\nu).$ Using the inequalities
\[ \widehat U_n(\nu^*)\le \widehat U_n(\widehat\nu_n),\quad U(\widehat\nu_n)\le U(\nu^*),\]
we get
\begin{align}
& U(\nu^*)-U(\widehat\nu_n)\le U(\nu^*)-\widehat U_n(\nu^*)+\widehat U_n(\widehat\nu_n)-U(\widehat\nu_n)\le U(\nu^*)-\widehat U_n(\nu^*)+\sup_{\nu\in\Delta} G_n(\nu), \label{1.3}\\
& \widehat U_n(\widehat\nu_n)-U(\nu^*)\le \widehat U_n(\widehat\nu_n)-U(\widehat\nu_n)\le\sup_{\nu\in\Delta} G_n(\nu). \label{1.4}
\end{align}
Note, that when $\nu_n$ is random, by $U(\nu_n)$ we mean  the conditional expectation:
\[ U(\nu_n)=\mathsf E\left( f(\nu_n,r_{n+1})|r_1,\dots,r_n\right)).\]
This quantity can be called the ``true utility'' of $\nu_n$ by analogy to the ``true risk'' in machine learning: see \cite{Shalev2014}. 

In learning theory the difference $U(\nu^*)-U(\widehat\nu_n)$ is called an estimation error: \cite{Shalev2014}. It describes the performance of the empirical utility maximizer $\widehat\nu_n$. The quantity $\widehat U_n(\widehat\nu_n)$ can be regarded as a statistical estimate of the true utility $U(\widehat\nu_n)$ of $\widehat\nu_n$. This estimate is always optimistically biased:
\[ \mathsf E U(\widehat\nu_n)\le U(\nu^*)=\mathsf E \widehat U_n(\nu^*)\le \mathsf E \widehat U_n(\widehat\nu_n).\]
The difference $\mathsf E \widehat U_n(\widehat\nu_n)- \mathsf E U(\widehat\nu_n)\ge 0$ is known as   optimizer's curse: \cite{Smith2006,Kuhn2019}.

We see that the key quantity is the supremum of the empirical process $G_n$. By the strong law of large numbers $G_n(\nu)\to 0$ a.s. for a fixed $\nu$. Moreover, since the function $\nu\mapsto u(\langle\nu,r\rangle)/u(r^*)$ is continuous and bounded, the convergence is uniform:
\[ \sup_{\nu\in\Delta}|G_n(\nu)|\to 0\ \textrm{a.s.,}\quad n\to\infty\]
by \cite[Theorem 7.53]{Shapiro2014}. From (\ref{1.3}), (\ref{1.4}) we see that
\[ U(\nu^*)\le \liminf_{n\to\infty} U(\widehat\nu_n), \quad \limsup_{n\to\infty} \widehat U_n(\widehat\nu_n)\le U(\nu^*).\]
The reverse inequalities $  U(\nu^*)\ge U(\widehat\nu_n)$,
\[ \liminf_{n\to\infty} \widehat U_n(\widehat\nu_n)\ge \liminf_{n\to\infty} \widehat U_n(\nu^*)=U(\nu^*)\]
imply that $\widehat U_n(\widehat\nu_n)\to U(\nu^*),$ $U(\widehat\nu_n)\to U(\nu^*)$, $n\to\infty$ a.s.
without further assumptions. Thus, the method of empirical utility maximization is consistent: see the definition in \cite[Chapter 3]{Vapnik1998}, where the convergence in probability is considered. In the next section  we  provide non-asymptotic bounds for $G_n$.

\section{Utility bounds}  \label{sec:3}
\setcounter{equation}{0}
Let us represent the supremum of the empirical process $G_n$ in the form
\[ \sup_{\nu\in\Delta} G_n(\nu)=\mathsf E \sup_{\nu\in\Delta} G_n(\nu)+\sup_{\nu\in\Delta} G_n(\nu)-\mathsf E \sup_{\nu\in\Delta} G_n(\nu).\]
Put $R_n=(r_1,\dots,r_n)$, $\Phi(R_n)=\sup_{\nu\in\Delta} G_n(\nu)$. We have
\begin{align*}
& |\Phi(r_1,\dots,\tilde r_k,\dots,r_n)-\Phi(r_1,\dots,r_k,\dots,r_n)|=\left|\sup_\nu\left(\frac{1}{m}\sum_{i\neq k}\frac{u(\langle\nu,r_i\rangle)}{u(r_i^*)}-U(\nu)+\frac{1}{m}\frac{u(\langle\nu,\tilde  r_k\rangle)}{u(\tilde r_k^*)} \right)\right.\\
& -\left.\sup_\nu\left(\frac{1}{m}\sum_{i\neq k}\frac{u(\langle\nu,r_i\rangle)}{u(r_i^*)}-U(\nu)+\frac{1}{m}\frac{u(\langle\nu,r_k\rangle)}{u(r_k^*)}\right)\right|\le\sup_\nu\left|\frac{1}{m}\frac{u(\langle\nu,\tilde r_k\rangle)}{u(\tilde r_k^*)}-\frac{1}{m}\frac{u(\langle\nu, r_k\rangle)}{u(r_k^*)}\right|\le\frac{1}{m}.
\end{align*}
By the McDiarmid concentration inequality (see \cite[Theorem D.8]{Mohri2018}) this bounded differences property implies that
\[ \mathsf P\left(\sup_\nu G_n(\nu)-\mathsf E \sup_\nu G_n(\nu)\ge\varepsilon\right)=
\mathsf P(\Phi(R_n)-\mathsf E\Phi(R_n) \ge\varepsilon)\le e^{-2m\varepsilon^2},
\]
or, equivalently,
\begin{equation} \label{2.1}
\mathsf P\left(\sup_\nu G_n(\nu)-\mathsf E \sup_\nu G_n(\nu)\ge\sqrt{\frac{1}{2n}\ln\frac{1}{\delta}}\right)\le\delta.
\end{equation}

For the difference $U(\nu^*)-\widehat U_n(\nu^*)$ we have a similar estimate:
\begin{equation} \label{2.2}
\mathsf P\left( U(\nu^*)-\widehat U_n(\nu^*)\ge\sqrt{\frac{1}{2n}\ln\frac{1}{\delta}}\right)\le\delta,
\end{equation}
which follows from the Hoeffding inequality \cite[Theorem D.2]{Mohri2018}: a  special case  of the  McDiarmid inequality.

Note, that to get the inequalities (\ref{2.1}), (\ref{2.2}) we need not impose any growth assumptions on $u$. This is an advantage of the relative utility. Let us formulate the obtained result more explicitly.

\begin{theorem} \label{th:1}
With probability at least $1-\delta$ we have
\begin{align}
U(\nu^*)-U(\widehat\nu_n) &\le \mathsf E \sup_{\nu\in\Delta} G_n(\nu)+\sqrt{\frac{2}{n}\ln\frac{2}{\delta}}, \label{2.3}\\ 
  \widehat U_n(\widehat\nu_n)-U(\widehat\nu_n) &\le \mathsf E \sup_{\nu\in\Delta} G_n(\nu)+\sqrt{\frac{1}{2n}\ln\frac{1}{\delta}}.  \label{2.4}
\end{align}
\end{theorem}

The distinction in constants in the right-hand sides of (\ref{2.3}), (\ref{2.4}) is due to the fact that we applied both inequalities (\ref{2.1}), (\ref{2.2}) to (\ref{1.3}) and only the first one to (\ref{1.4}). In the first case the following argumentation is used: if
\[ \mathsf P\left(\xi_i\ge \sqrt{\frac{1}{2n}\ln\frac{1}{\delta}}\right)\le \delta,\quad i=1,2, \]
then
\[ \mathsf P\left(\xi_1+\xi_2\ge 2\sqrt{\frac{1}{2n}\ln\frac{2}{\delta}}\right)\le\sum_{i=1}^2 \mathsf P\left(\xi_i\ge \sqrt{\frac{1}{2n}\ln\frac{1}{\delta/2}}\right)\le\delta.\]

Theorem \ref{th:2} contains the main result of the paper: the upper bounds for $\mathsf E \sup_{\nu\in\Delta} G_n(\nu)$. 

\begin{theorem} \label{th:2}
Assume that the utility function $u$ is uniformly H\"{o}lder continuous on $(0,\infty)$:
\begin{equation} \label{2.5}
 |u(x)-u(y)|\le K|x-y|^\alpha
\end{equation} 
with some $\alpha\in (0,1]$, $K>0$. Assume further that 
\begin{equation} \label{2.6}
 A:=\sup_{x>0}\frac{x^\alpha}{u(x)}<\infty. 
\end{equation} 
 Then 
 \begin{align}
\mathsf E \sup_{\nu\in\Delta} G_n(\nu)&\le  2AK \sqrt{\frac{2\ln d}{n}}, \quad \alpha =1, \label{2.7}\\
\mathsf E \sup_{\nu\in\Delta} G_n(\nu)&\le CAK \sqrt{\frac{d-1}{\alpha n}},\quad \alpha \in (0,1), \label{2.8}
\end{align}
where $C>0$ is an absolute constant. 
\end{theorem}

\emph{Proof}. Let $\varepsilon_i$, $i=1,\dots,n$ be independent Rademacher random variables: $\mathsf P(\varepsilon_i=1)=\mathsf P(\varepsilon_i=-1)=1/2$,  which are also independent from $r_1,\dots, r_n$. Consider the empirical Rademacher complexity (see , e.g.,  \cite{Mohri2018})
\[ \widehat{\mathcal R}(\mathcal F\circ R_n)=\frac{1}{n}\mathsf E\left(\sup_{\nu\in\Delta}\sum_{i=1}^n\varepsilon_i\frac{u(\langle \nu,r_i\rangle)}{u(r_i^*)} \biggl | R_n\right) \]
of the set of functions $\mathcal F=\{r\mapsto  u(\langle \nu,r\rangle)/u(r^*):\nu\in\Delta\}$ with respect to the random sequence $R_n=(r_1,\dots,r_n)$.  In fact we compute the Rademacher complexity of the following set of $n$-dimensional vectors:
\[ \mathcal F\circ R_n:=\left\{\left(\frac{u(\langle \nu,r_1\rangle)}{u(r_1^*)},\dots,\frac{u(\langle \nu,r_n\rangle}{u(r_n^*)}\right): \nu\in\Delta \right\}. \]
For clarity recall (see \cite{Shalev2014}) that the Rademacher complexity of a set $C\subset\mathbb R^n$ is defined by the formula 
\begin{align} \label{2.9}
 \widehat{\mathcal R}(C)=\frac{1}{n}\mathsf E \sup_{a\in C} \sum_{i=1}^n\varepsilon_i a_i.
\end{align}

Let us consider the case $\alpha=1$. The symmetrization argument (\cite[Lemma 7.4]{vanHandel2016}) gives the bound
\begin{equation} \label{2.10}
 \mathsf E \sup_{\nu\in\Delta} G_n(\nu)\le 2 \mathsf E\widehat{\mathcal R}(\mathcal F\circ R_n).
\end{equation} 
For  $\Psi(x,r)=u(x)/u(r^*)$, $r^*=\max_{1\le i\le d} r^i$ we have
\[  |\Psi(x,r)-\Psi(y,r)|\le \frac{K}{u(r^*)}|x-y|.\]
Literally following the proof of Talagrand's contraction lemma, given in \cite[Lemma 5.7]{Mohri2018}, we get the inequality
\begin{align}
\widehat{\mathcal R}(\mathcal F\circ R_n) &=
\frac{1}{n}\mathsf E\left(\sup_{\nu\in\Delta}\sum_{i=1}^n\varepsilon_i \Psi(\langle \nu,r_i\rangle,r_i)
\biggr| R_n\right)\le \frac{K }{n}\mathsf E\left(\sup_{\nu\in\Delta}\sum_{i=1}^n\varepsilon_i \frac{\langle \nu,r_i\rangle}{r_i^*}
\biggr| R_n\right)\nonumber\\
&= K \widehat{\mathcal R}(\mathcal H\circ R_n), \quad \mathcal H:=\{r\mapsto  \langle \nu,r\rangle/r^*:\nu\in\Delta\}. \label{2.11}
\end{align}
Note, that the only difference with the Talagrand contraction lemma is that the Lipschitz constant for $x\mapsto \Psi(x,r)$ depends on $r$.

The Rademacher complexity of the set $\mathcal H$ equals to the Rademacher complexity of its extreme points (as follows from \cite[Lemma 26.7]{Shalev2014}), corresponding to the vectors of the standard basis: $\nu\in\{e_1,\dots,e_d\}$, $e_i=(\delta_{ij})_{j=1}^d$, where $\delta_{ij}$ is Kronecker symbol. Thus,
\begin{equation} \label{2.12}
\widehat{\mathcal R}(\mathcal H\circ R_n)=\widehat{\mathcal R}\left(\frac{r^1}{u(r^*)},\dots, \frac{r^d}{u(r^*)}\right).
\end{equation}
Here $r^j/u(r^*)=(r^j_1/u(r^*_1),\dots, (r^j_n/u(r^*_n))\in\mathbb R^n$ are the normalized trajectories of the returns, and the right-hand side of (\ref{2.12}) is computed in accordance with (\ref{2.9}).  The Rademacher complexity of a finite set of vectors can be estimated by Massart's lemma (see \cite[Theorem 3.7]{Mohri2018}). Applying this lemma to the right-hand side of (\ref{2.12}), we get the inequality 
\begin{equation} \label{2.13}
 \widehat{\mathcal R}\left(\frac{r^1}{u(r^*)},\dots, \frac{r^d}{u(r^*)}\right) \le \frac{A}{\sqrt n}\sqrt{2\ln d}, 
\end{equation}
since by (\ref{2.6}),
\[ \|r^j/u(r^*)\|_2 =\sqrt{\sum_{k=1}^n \left(\frac{r^j_k}{u(r^*_k)}\right)^2}\le A\sqrt n,\]
where $\|a\|_2=\sqrt{\sum_{i=1}^n a_i^2}$ is the $l_2$-norm.
The inequality (\ref{2.7}) now follows from (\ref{2.10}) -- (\ref{2.13}).

In the case $\alpha<1$ first note that for fixed $R_n$ the process
\[ Z_n(\nu)=\frac{1}{n}\sum_{k=1}^n\varepsilon_k \frac{u(\langle \nu,r_k\rangle)}{u(r_k^*)} \]
is subgaussian (see \cite[Definition 5.20]{vanHandel2016}) with respect to the data dependent pseudometric 
\[ \rho(\nu,\nu')=\frac{1}{n}\left(\sum_{k=1}^n \left(\frac{u(\langle\nu,r_k\rangle)}{u(r_k^*)}-\frac{u(\langle\nu',r_k\rangle)}{u(r_k^*)}\right)^2 \right)^{1/2},\]
defined on $\Delta$. That is, 
\begin{align*}
\mathsf E\left(e ^{\lambda (Z_n(\nu)-Z_n(\nu'))}\biggr|R_n\right)=\prod_{i=1}^n \mathsf E\left[\exp\left(\frac{\lambda}{n}\varepsilon_i \frac{u(\langle \nu,r_k\rangle)-u(\langle \nu',r_k\rangle)}{u(r_k^*)} \right)\biggr|R_n\right]\le e^{\lambda^2 \rho^2(\nu,\nu')/2}.
\end{align*}
Here we used an elementary inequality $\mathsf E e^{\lambda\varepsilon_i a}\le e^{\lambda^2 a^2/2}$: \cite[Example 2.3]{Wainwright2019}.

A set $N\subset\Delta$ is called $\epsilon$-dispersed if $\rho(\nu,\nu')\ge\epsilon$ for $\nu,\nu'\in N$ with $\nu\neq\nu'$. Let $D(\Delta,\rho,\epsilon)$ be the $\epsilon$-packing number of $(\Delta,\rho)$:
\[ D(\Delta,\rho,\epsilon)=\sup\{|N|: N \textrm{ is an } \epsilon\textrm{-dispersed} \}.\]
Here $|N|$ is the cardinality of $N$. The conditional expectation of the supremum of $Z_n$ is bounded by the Dudley entropy integral (\cite[Corollary 13.2]{Boucheron2013}):
\begin{equation} \label{2.14}
\widehat{\mathcal R}(\mathcal F\circ R_n)=\mathsf E\left(\sup_{\nu\in\Delta} Z_n(\nu)|R_n\right)\le 12\int_0^{d/2}\sqrt{\ln D(\Delta,\rho,\epsilon)}\,d\epsilon,
\end{equation}
where $d$ is the diameter of $\Delta$. 

Conditions (\ref{2.5}), (\ref{2.6}) imply that
\begin{align}
\rho(\nu,\nu') &\le \frac{K}{n}\left(\sum_{k=1}^n \frac{|\langle\nu-\nu',r_k\rangle|^{2\alpha}}{u^2(r_k^*)}\right)^{1/2}\le \frac{K}{n}\left(\sum_{k=1}^n \frac{(r_k^*)^{2\alpha}\|\nu-\nu'\|_1^{2\alpha}}{u^2(r_k^*)}\right)^{1/2} \nonumber \\
&\le  \frac{KA}{\sqrt n} \|\nu-\nu'\|_{1}^{\alpha},\quad \label{2.15}
\end{align}
where $\|a\|_{1}=\sum_{j=1}^d |a_j|$ is the the $l_1$-norm.
For the $\epsilon$-packing number of $\Delta$ with the metric, induced by $\|\cdot\|_1$, we have the inequality 
$D(\Delta,\|\cdot\|_1,\epsilon)\le \left(5/\epsilon\right)^{d-1}$
(see \cite[Proposition C.1]{Ghosal2017}). From (\ref{2.15}) it follows that if $\rho(\nu,\nu')\ge\epsilon$ then
\[ \|\nu-\nu'\|_1\ge \left(\frac{\sqrt n \varepsilon}{KA}\right)^{1/\alpha}.  \] 
Hence,
\begin{equation} \label{2.16}
 D(\Delta,\rho,\epsilon)\le D\left(\Delta,\|\cdot\|_1,\left(\frac{\sqrt{n}\epsilon}{KA} \right)^{1/\alpha}\right)\le 5^{d-1}\left(\frac{KA}{\sqrt{n}\epsilon}\right)^{(d-1)/\alpha}.
\end{equation} 
Furthermore, by (\ref{2.15}) the diameter of $\Delta$ with respect to $\rho$ is estimated as
\begin{equation} \label{2.17}
 d\le 2^\alpha\frac{KA }{\sqrt n},
\end{equation} 
since $\|\nu-\nu' \|_1\le \|\nu\|_1+ \|\nu' \|_1\le 2$.
Let us substitute the estimates (\ref{2.16}), (\ref{2.17}) into (\ref{2.14}), and perform the change of variables $z=\sqrt n\varepsilon/(2^{\alpha-1} KA)$:
\begin{align*}
\widehat{\mathcal R}(\mathcal F\circ S_n) &\le 12\int_0^{2^{\alpha-1} KA/\sqrt n} \sqrt{\ln \left(5^{d-1} \left(\frac{KA}{\sqrt{n}\epsilon}\right)^{(d-1)/\alpha}\right)}\,d\epsilon\\
&=12\sqrt\frac{d-1}{\alpha} \int_0^{2^{\alpha-1} KA/\sqrt n} \sqrt{\ln \left(5^\alpha \frac{KA}{\sqrt{n}\epsilon}\right)}\,d\epsilon\\
&=12\sqrt\frac{d-1}{\alpha}\frac{2^{\alpha-1} KA}{\sqrt n}\int_0^1 \sqrt{\ln\frac{5^\alpha}{2^{\alpha-1} z}}\,dz \le C_1  KA \sqrt\frac{d-1}{\alpha n},\\
C_1 &=12\int_0^1\sqrt{\ln\frac{5}{z}}\,dz.
\end{align*}
Together with (\ref{2.10}) this completes the proof ($C=2C_1$). \qed

In a most natural way condition (\ref{2.6}) is satisfied by the power utility function $u(x)=x^\alpha$, $\alpha\in (0,1]$. This function also satisfies (\ref{2.5}) with $K=1$, as easily follows from the inequality (\cite[Appendix A, Lemma 5.1]{Gut2013})
\[ (x+y)^\alpha \le x^\alpha+ y^\alpha,\quad x,y>0. \]

For $u(x)=x^\alpha$ the problem (\ref{1.1}) reduces to the optimization of the ordinary power utility function after the price normalization:
\[ U(\nu)=\mathsf E \langle\nu,r_{n+1}/r_{n+1}^*\rangle^\alpha.\]
The power utility is natural in one more respect: the relative utility (\ref{2.1}) in this case is independent of investor's wealth $x$:
\[ \mathsf E \frac{u(x\langle\nu,r_{n+1}\rangle)}{u\left(x r_{n+1}^* \right)}=\mathsf E \langle\nu,r_{n+1}/r_{n+1}^*\rangle^\alpha. \]
This means that one can consider the problems (\ref{1.1}), (\ref{1.2}) dynamically in an online manner. At each step the investor will act myopically similar to the case of the ordinary logarithmic utility.

\begin{remark} Under additional assumptions condition (\ref{2.6}) on the utility function can be relaxed. In fact we need only the upper bound for $r_k^*/u(r_k^*)$. Thus, if there exists a riskless asset (cash) with $r_k=1$, then the supremum in (\ref{2.6}) can be taken over $[1,\infty)$. Furthermore, if the returns are bounded, then the supremum can be taken over a finite interval. In this case usually it is enough to consider the Lipschitz case $\alpha=1$.
\end{remark}

\begin{remark}
Theorems \ref{th:1}, \ref{th:2} give high probability error bounds. From (\ref{1.3}), (\ref{1.4}) it follows that
\begin{align*}
\max\{U(\nu^*)- \mathsf E U(\widehat\nu_n), \mathsf E (\widehat U_n(\widehat\nu_n)-U(\widehat\nu_n)) \} \le \mathsf E \sup_{\nu\in\Delta} G_n(\nu),
\end{align*}
Thus, Theorem 2 provides also error bounds in expectation. 
\end{remark}

\begin{remark}
The obtained error bounds are of order $n^{-1/2}$.  In general the main assumption, which allows to obtain $O(1/n)$ bounds, is the strong concavity of $U$: \cite{Shalev2010,Rakhlin2011}. However, such assumption requires additional conditions on the returns $r_i$, which we want to avoid in the present paper.
\end{remark}

\section{Stochastic exponentiated gradient algorithm}  \label{sec:4}
\setcounter{equation}{0}
In this section we additionally assume that the utility function $u$ is concave. Recall that the subdifferential of $-u$ at any point $y\in (0,\infty)$ is an interval:
\[ \partial(-u)(y)=[ -D_-u(y),-D_+u(y)],\]
where $D_ -u(y)$ and $D_+u(y)$ are the left and right derivatives: see \cite[Chap.\,I]{Hiriart1993}. We have  $D_ -u(y)\ge D_+u(y)\ge 0$, as $u$ is non-decreasing. 

We use the exponentiated gradient (EG) algorithm of \cite{Kivinen1997} to solve the empirical utility maximization problem (\ref{1.2}). Consider the empirical distribution generated by the sample $(r_1,\dots,r_n)$, and a random variable $\widehat r$ with this distribution:
\[ \widehat{\mathsf P}(\widehat r =r_k)=\frac{1}{n},\quad k=1,\dots, n.\]
Put 
\[\underline r_n=\min_{1\le k\le n} \min_{1\le i\le d} r_k^i,\quad \overline r_n=\max_{1\le k\le n}\max_{1\le i\le d} r_k^i\] 
and consider the convex functions 
\[ \nu\mapsto f_j(\nu)=1-\frac{u(\langle\nu,\widehat  r_j\rangle)}{u(\widehat r_j^*)}:\Delta\mapsto [0,1]. \]
From the description of their subdifferentials:
\[ \partial f_j(\nu)=\left\{\frac{\gamma}{u(\widehat r_j^*)}\widehat r_j:\gamma\in [-D_- u(\langle\nu,\widehat  r_j\rangle),- D_+ u(\langle\nu,\widehat  r_j\rangle)]\right\} \]
and the inequalities $0<\underline r_n\le \langle\nu,\widehat  r_j\rangle$, $j=1,\dots,n,$ we see that the absolute values of the subgradient components are bounded by the constant
\[ L_n=D_- u(\underline r_n)\cdot\max_{\underline r_n\le x\le\overline r_n} \frac{x}{u(x)}=
D_- u(\underline r_n)\cdot \frac{\overline r_n}{u(\overline r_n)}.
\]
Indeed, $u(x)/x$ is non-increasing: \cite[Proposition 1.1.4]{Hiriart1993}, and the subdifferential mapping is monotone:
\[ \gamma_1\le\gamma_2 \quad \textrm{whenever}\quad \gamma_i\in\partial(-u)(y_i),\quad 0<y_1<y_2,\]
see \cite[Theorem 4.2.1]{Hiriart1993}. It follows that the functions $f_j$ are $L_n$-Lipschitz with respect to $l_1$-norm: see \cite[Lemma 2.6]{Shalev2012}.

Apply the exponentiated gradient algorithm to $f_1, \dots, f_m$:
\begin{align}
\nu_0^i &=1/d,\quad i=1,\dots,d, \label{3.1}\\
a_j^i &=\nu_{j-1}^i \exp\left(\eta\frac{D_-u(\langle\nu_{j-1},\widehat r_j\rangle)}{u(\widehat r_j^*)} \widehat r_j^i \right),\quad \nu_j^i=\frac{a_j^i}{\sum_{l=1}^d a_j^l}, \label{3.2}
\end{align}
$i=1,\dots,d$, $j=1,\dots,m-1$, where $\eta>0$ is a parameter. Note that,
\[-\frac{D_-u(\langle\nu_{j-1},\widehat r_j\rangle)}{u(\widehat r_j^*)}\widehat r_j\in  \partial f_j(\nu).\]

For a moment assume that $\widehat r_j\in (0,\infty)^d$ is an arbitrary sequence. The basic problem of the
online convex optimization theory is to find a sequence $\nu_0,  \dots, \nu_{m-1}$ such that $\nu_{j-1}$ does not depend on $f_j,\dots, f_m$ and the regret
\[ \textrm{Regret}_m(\nu)=\sum_{j=1}^m f_j(\nu_{j-1})-\sum_{j=1}^m f_j(\nu)=\sum_{j=1}^m \frac{u(\langle\nu,\widehat  r_j\rangle)}{u(\widehat r_j^*)}-\sum_{j=1}^m \frac{u(\langle\nu_{j-1},\widehat  r_j\rangle)}{u(\widehat r_j^*)}\]
is small uniformly over $\nu\in\Delta$. It is well known that the EG algorithm with $\eta=\sqrt{\frac{\ln d}{m}}\frac{1}{L_n}$ ensures the estimate
\begin{equation} \label{3.3}
\textrm{Regret}_m(\nu)\le 2 L_n \sqrt{m}\sqrt{\ln d},
\end{equation}
see \cite[Corollary 2.14]{Shalev2012} (a constant is corrected).

For an i.i.d. random sequence $\widehat r_j$ we can apply to (\ref{3.1}), (\ref{3.2}) the online-to-batch conversion scheme: \cite[Chap. 5]{Shalev2012}. In this case it is natural to call (\ref{3.1}), (\ref{3.2}) the stochastic exponentiated gradient (SEG) algorithm. Denote by $\widehat{\mathsf E}$ is the expectation with respect to the empirical distribution of $r_1,\dots, r_n$. For any fixed $\nu$,
\begin{equation} \label{3.4}
 \widehat{\mathsf E} \frac{u(\langle\nu,\widehat r_j\rangle)}{u(\widehat r_j^*)}=\frac{1}{n}\sum_{k=1}^n\frac{u(\langle\nu, r_k\rangle)}{u(r_k^*)}=\widehat U_n(\nu).
\end{equation}
Furthermore, since $\nu_{j-1}$ is $\sigma(\widehat r_1,\dots,\widehat r_{j-1})$-measurable, we have
\[\widehat{\mathsf E}\frac{u(\langle\nu_{j-1},\widehat r_j\rangle)}{u(\widehat r_j^*)} =
 \widehat{\mathsf E} \widehat{\mathsf E}\left(\frac{u(\langle\nu_{j-1},\widehat r_j\rangle)}{u(\widehat r_j^*)}\biggl|\widehat r_1,\dots,\widehat r_{j-1}\right)= \widehat{\mathsf E}\frac{1}{n}\sum_{k=1}^n \frac{u(\langle\nu_{j-1},r_k\rangle)}{u(r_k^*)}, \]
\begin{align}  
\frac{1}{m}\widehat{\mathsf E}\sum_{j=1}^m\frac{u(\langle\nu_{j-1},r_j\rangle)}{u(r_j^*)} &=\frac{1}{m}\sum_{j=1}^m  \widehat{\mathsf E}\frac{1}{n}\sum_{k=1}^n \frac{u(\langle\nu_{j-1},r_k\rangle)}{u(r_k^*)}
=\frac{1}{n}\sum_{k=1}^n\widehat{\mathsf E}\frac{1}{m}\sum_{j=1}^m\frac{u(\langle\nu_{j-1},r_k\rangle)}{u(r_k^*)}\nonumber\\
& \le
\widehat{\mathsf E}\frac{1}{n}\sum_{k=1}^n \frac{u(\langle\overline\nu_m,r_k\rangle)}{u(r_k^*)}=\widehat{\mathsf E}\widehat U_n(\overline\nu_m),  \label{3.5}
\end{align}
where
\begin{equation} \label{3.6}
\overline\nu_m=\frac{1}{m}\sum_{j=0}^{m-1} \nu_j. 
\end{equation}
In these calculations $r_1,\dots,r_n$ are regarded as constants. Note that $\nu_j$, $\overline\nu_m$ depend also on $n$, but we suppress this dependence in the notation.

From (\ref{3.3}) -- (\ref{3.5}) we get
\begin{align*}
2 L_n \sqrt\frac{\ln d}{m}\ge\widehat{\mathsf E}\frac{\textrm{Regret}_m(\nu)}{m}&=\frac{1}{m}\widehat{\mathsf E}\sum_{j=1}^m \left(\frac{u(\langle\nu,\widehat r_j\rangle)}{u(\widehat r_j^*)}-\frac{u(\langle\nu_{j-1},\widehat r_j\rangle)}{u(\widehat r_j^*)} \right)\ge \widehat U_n(\nu)-\widehat{\mathsf E}\widehat U_n(\overline\nu_m).
\end{align*}
In particular, for an empirical utility maximizer $\widehat\nu_n$,
\begin{equation} \label{3.7}
\widehat U_n(\widehat\nu_n)\le  \widehat{\mathsf E}\widehat U_n(\overline\nu_m)+2 L_n \sqrt{\frac{\ln d}{m}}\le \widehat U_n(\overline\nu_m)+\sqrt{\frac{1}{2 n}\ln\frac{1}{\delta}}+2 L_n \sqrt{\frac{\ln d}{m}}
\end{equation}
with probability at least $1-\delta$ by Hoeffding's inequality (\cite[Theorem D.2]{Mohri2018}):
\begin{align*}
\widehat{\mathsf P}( \widehat{\mathsf E}\widehat U_n(\overline\nu_m)-\widehat U_n(\overline\nu_m)\ge\varepsilon )=\widehat{\mathsf P}\left(\frac{1}{n}\sum_{k=1}^n \frac{u(\langle\overline\nu_m,r_k\rangle)}{u(r_k^*)}-\widehat{\mathsf E}\frac{1}{n}\sum_{k=1}^n \frac{u(\langle\overline\nu_m,r_k\rangle)}{u(r_k^*)}\ge\varepsilon \right)\le e^{-2\varepsilon^2 n }
\end{align*}
with $\varepsilon=\sqrt{\frac{1}{2n}\ln\frac{1}{\delta}}$.

We now able to provide for $\overline\nu_m$ an analog of inequality (\ref{2.3}):
\begin{align*}
U(\nu^*)-U(\overline\nu_m) &=U(\nu^*)-\widehat U_n(\nu^*)+\widehat U_n(\nu^*)-\widehat U_n(\nu_n)+\widehat U_n(\nu_n)-\widehat U_n(\overline\nu_m)+\widehat U_n(\overline\nu_m)-U(\overline\nu_m)\\
& \le  (U(\nu^*)-\widehat U_n(\nu^*))+(\widehat U_n(\nu_n)-\widehat U_n(\overline\nu_m))+\sup_{\nu\in\Delta} G_n(\nu).
\end{align*} 
Applying (\ref{2.2}), (\ref{3.7}) and (\ref{2.1}) respectively to the tree terms in the right-hand side, we get the following result.
\begin{theorem} \label{th:3}
Assume that the function $u$ is concave. Then for the average portfolio (\ref{3.6}), produced by the SEG algorithm (\ref{3.1}), (\ref{3.2}), with probability at least $1-3\delta$ the following estimate holds true:
\begin{align*}
U(\nu^*)-U(\overline\nu_m) \le   \mathsf E \sup_{\nu\in\Delta} G_n(\nu)+3\sqrt{\frac{1}{2n}\ln\frac{1}{\delta}}+2 L_n \sqrt{\frac{\ln d}{m}}.
\end{align*} 
\end{theorem}

Certainly, the estimates of Theorem \ref{th:2} still can be applied to $\mathsf E \sup_{\nu\in\Delta} G_n(\nu)$. Thus, Theorem \ref{th:3} gives a high-probability bound for the estimation error of the stochastic exponentiated gradient algorithm. The value of $m$ can be taken sufficiently large to get  for the estimation error of $\overline\nu_m$ the bound of the same order as for the exact empirical utility maximizer $\widehat \nu_n$. The mentioned value of $m$ is data dependent, since the Lipschitz constant $L_n$ depends on the  returns $(r_1,\dots,r_n)$. Note, that we need no new data to generate an arbitrary large sample $\widehat r_1,\dots,\widehat r_m$ used in the SEG algorithm. 

\section{Power utility: the case of one risky asset} \label{sec:5}
\setcounter{equation}{0}
Consider the case $d=2$. In this section we will put upper indexes in brackets.  Assume that the investor can keep money in cash: $r^{(1)}_t=1$, or invest in a risky asset, whose daily returns are log-normal and follow the discrete-time Black-Scholes model:
\begin{equation} \label{4.1}
  r_k^{(2)}=\exp\left(\frac{\mu -\sigma^2/2}{T} +\frac{\sigma}{\sqrt{T}} Z_k\right),\quad \quad k=1,\dots,n.  
\end{equation}  
Here $T=252$ is the number of trading days in a year; $Z_k$ are independent standard normal variables: $Z_k\sim N(0,1)$; $n$ is the sample size, which we assume to be multiple of $T$.  Put $\mu=0.15$, which corresponds to 
\[\mathsf E \prod_{k=1}^T r_k^{(2)}=e^{\mu}\approx 1.162\]
annual expected return for the risky asset, and $\sigma=0.45$. We have 
\[ \ln r_k^{(2)}\sim N\left(\frac{\mu -\sigma^2/2}{T},\frac{\sigma}{\sqrt{T}}\right)=N(1.93\cdot 10^{-4},2.83\cdot 10^{-2}).\]

In this section we assume that $u(x)=x^\alpha$, $\alpha\in (0,1]$. The the relative empirical utility maximization problem (\ref{1.2}) takes the form
\begin{equation} \label{4.2}
\psi(\nu^{(2)})=\frac{1}{n}\sum_{k=1}^n  \langle\nu,r_k/r_k^*\rangle^\alpha=\frac{1}{n}\sum_{k=1}^n  \left(\frac{1}{\max\{1,r_k^{(2)}\}}+\frac{r_k^{(2)}-1}{\max\{1,r_k^{(2)}\}}\nu^{(2)}\right)^\alpha\to\max_{\nu^{(2)}\in [0,1]}.
\end{equation}
For comparison consider also the ordinary empirical utility:
\begin{equation} \label{4.3}
 \varphi(\nu^{(2)})=\frac{1}{n}\sum_{k=1}^n \langle\nu,r_k\rangle^\alpha=\frac{1}{n}\sum_{k=1}^n \left(1+(r_k^{(2)}-1)\nu^{(2)}\right)^\alpha\to\max_{\nu^{(2)}\in [0,1]}. 
\end{equation}

For a large $n=T\cdot 10^3=2.52\cdot 10^5$ we applied to $\varphi'(\nu_2), \psi'(\nu_2)$ the bisection method {\tt optimize.bisect} from the module {\tt scipy} (Python) with the default tolerance parameter. The results, averaged over 100 realizations of $(r_k^{(2)})_{k=1}^n$, are presented in Table \ref{tab:1}.
\begin{table}
\caption{Average optimal weight $\nu^{(2)}$ of the risky asset}
\label{tab:1}
\begin{tabular}{ c c  c  c  c  c  c  c  c  c }
\hline\noalign{\smallskip}
 $\alpha$ & 0.001 & 0.01 & 0.1 & 0.2 & 0.3 & 0.5 & 0.75 & 0.9 \\
 \hline
 \makecell{Ordinary power  \\  utility, $\varphi$} & 0.7380 & 0.7448 & 0.8188 &0.9118 & 0.9775 & 1 & 1 & 1   \\ 
 \hline
\makecell{Relative power\\ utility, $\psi$} & 0.7376 & 0.7397 & 0.7637& 0.7961 & 0.8367 & 0.9245 & 0.9909 &1 \\ 
  \hline
\end{tabular}
\end{table}

We see that the relative utility makes the investor more risk averse. This property can be easily explained. Instead of the power utility function consider a differentiable increasing concave function $u$.  Without loss of generality, we can assume that $u(1)=1$. For the expected utilities, corresponding to (\ref{4.2}), (\ref{4.3}),  we have
\begin{align*}
&\psi'(\nu^{(2)}):=\frac{\partial U(\nu)}{\partial\nu^{(2)}}=\mathsf E\left(\frac{u'(1+(r^{(2)}-1)\nu^{(2)})}{u(\max\{1, r^{(2)}\})}(r^{(2)}-1) \right)\\
&=\mathsf E\left(u'(1+(r^{(2)}-1)\nu^{(2)})(r^{(2)}-1) I_{\{r^{(2)}\le 1\}}\right)
+\mathsf E\left(\frac{u'(1+(r^{(2)}-1)\nu^{(2)})}{u(r^{(2)})}(r^{(2)}-1) I_{\{r^{(2)}>1\}}\right) \\ 
&\le\mathsf E\left(u'(1+(r^{(2)}-1)\nu^{(2)})(r^{(2)}-1)\right)=\frac{\partial \widetilde U(\nu)}{\partial\nu^{(2)}}=:\varphi'(\nu^{(2)}),
\end{align*}
where $\widetilde U(\nu)=\mathsf E u(\langle\nu,r\rangle)$ is the ordinary expected utility. The functions $\psi'$, $\varphi'$ are decreasing. It follows that the zero  of $\psi'$ is smaller than the zero of $\varphi'$ (for simplicity we assume that a zero is unique). A similar argumentation works for the empirical utilities.

However, in the next section we will see that the discussed property is not universal. In a model with several risky assets the optimal portfolio, corresponding to the relative power utility, can be more risky, than for the ordinary utility. 

Next we argue that if the price of a risky asset follows the Black-Scholes model, neither $10$ nor $100$ years are enough to make any reliable conclusions concerning the optimal value $\nu^{(*,2)}$ on the basis of daily historical prices. 

\begin{figure}
\centering
\includegraphics[scale=0.95]{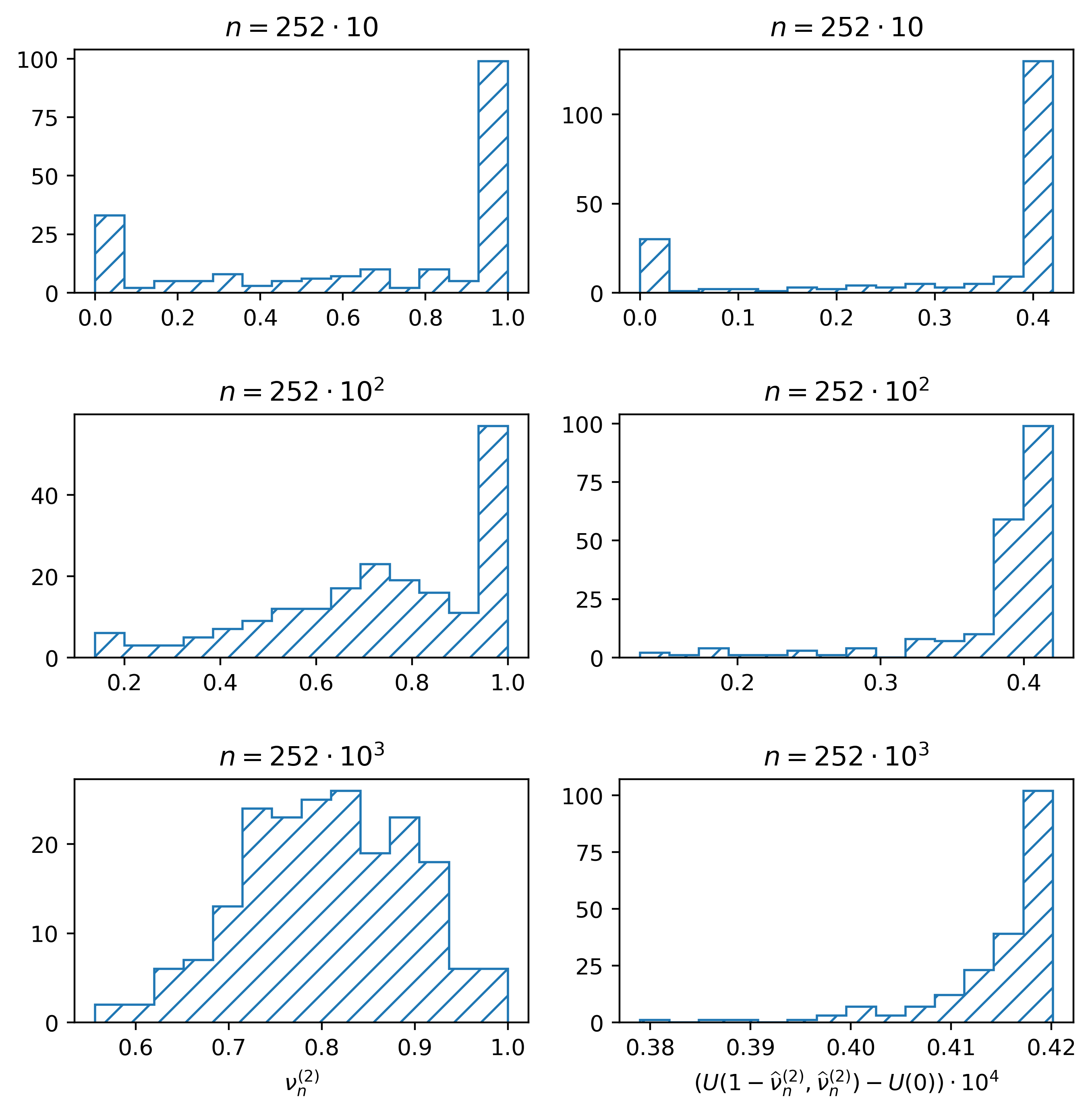}
\caption{Histograms of optimal weight $\widehat\nu^{(2)}_n$ of the risky asset (left panels) and of
linearly transformed true utility $(U(\widehat\nu_n)-U(w_0))\cdot 10^4$, $w_0=(1,0)$ (right panels) for 200 realizations of daily returns $(r_k^{(2)})_{k=1}^n$ for  $n=252\cdot 10^k$, $k=1,2,3$. The case of relative power utility with $\alpha=0.2$.}
 \label{fig:1}
\end{figure}

For $\alpha=0.2$ in the left panels of Fig.\,\ref{fig:1} we show the histograms of the optimal weight $\widehat\nu^{(2)}_n$ of the risky asset for 200 realizations of daily returns $(r_k^{(2)})_{k=1}^n$, where  $n=252\cdot 10^k$, $k=1,2,3$.  To estimate the true utility $U(\nu)$ of $\widehat\nu$  we used  the empirical  mean $\widehat U_N(\nu)$ with very large $N=10^7$. The histogram of linearly transformed true utilities $(U(\widehat\nu)-U(w_0))\cdot 10^4$, $w_0=(1,0)$ are shown in the right panels in Fig.\,\ref{fig:1}. In the same way we obtained the estimates of the optimal weight of the risky asset: $\nu^{*,2}\approx 0.81$, and its utility
\begin{equation} \label{4.4}
 (U(\widehat\nu^*)-U(w_0))\cdot 10^4\approx 0.42.
\end{equation} 

We see that optimal portfolio weights very slowly concentrate near the optimal value. In particular for 
$n=252\cdot 10$ in most cases $\widehat\nu_n^{(2)}$ simply takes the extreme values 0 and 1. Only for  $n=252\cdot 10^3$ the largest peak is near the optimum. But even in this case it is blurred. Note, however, that the true utilities of $\widehat\nu_n^{(2)}$ demonstrate somewhat better concentration near the optimum (\ref{4.4}). These conclusions are not specific for the relative power utility or for a specific value of $\alpha$. For  for other values of $\alpha$, and for the ordinary power or logarithmic utilities the results will be similar. 

Note that the slow concentration phenomenon (which is related to the fragility of SAA in portfolio optimization:  \cite{Ban2018}) does not contradict Theorems \ref{th:1}, \ref{th:2}. Roughly speaking, these  theorems give the estimate
\[ U(\nu^*)-U(w_0)\le U(\widehat\nu_n)-U(w_0)+O\left(\frac{1}{\sqrt n}\right) \]
with high probability. From (\ref{4.4}) it follows that we need $n$ at least of order $10^8$ to get a nontrivial lower bound for $U(\widehat\nu_n)-U(w_0)$.

\section{Experiments with NYSE data}  \label{sec:6}
\setcounter{equation}{0}
We considered two datasets, containing daily stock returns form the New-York Stock Exchange (NYSE): 
\begin{itemize}
\item NYSE$_1$: Contains 5651 daily returns of 36 stocks for the period ending in 1984, 
\item NYSE$_2$:  Contains 11178 daily returns of 19 stocks for the period ending in 2006.
\end{itemize}
Both datasets were taken from 
\url{http://www.cs.bme.hu/~oti/portfolio/data.html}. NYSE$_1$ is a classical dataset, considered in many papers, starting from \cite{Cover1991} (see the references in \cite{Gyorfi2012,Gyorfi2017}). NYSE$_2$ was first analized in \cite{Gyorfi2012}, where the authors also proposed a simple greedy algorithm for the empirical logarithmic utility maximization: 
\[\frac{1}{n}\sum_{k=1}^n \ln \langle\nu,r_k\rangle\to\max_{\nu\in\Delta}.\]

In this paper we are interested in an application of the exponentited gradient (EG) algorithm. Note that already in \cite{Helmbold1998} this algorithm was applied to the NYSE$_1$ dataset and the logarithmic utility. However, our goal here is different: we want to solve the problem (\ref{1.2}). Unfortunately we were unable to do this using the algorithm in the form (\ref{3.1}), (\ref{3.2}) or with time-varying learning rate $\eta$ (e.g., applying the doubling trick: see \cite{Shalev2012}). So, we propose its modification: the greedy doubly stochastic exponentiated gradient (GDSEG) algorithm. For clarity we present its pseudocode for the power utility $u(x)=x^\alpha$.

\begin{algorithm}
\caption{Greedy doubly stochastic exponentiated gradient algorithm  (GDSEG) for the power utility}
\begin{algorithmic}[1]
\renewcommand{\algorithmicrequire}{\textbf{Input:}}
\Require $\overline\eta>0$: an upper bound  for learning rate; $\texttt{n\_attempts}$: an upper bound  for the number of attempts to improve a current portfolio; \texttt{threshold}: an improvement threshold; $\{r_k^i: k\in\{1,\dots,n\}, i\in\{1,\dots,d\}\}$: an array of daily returns; $\alpha\in (0,1]$
\State $\nu^i:=1/d$, $i=1,\dots,d$
\If {the relative utility is considered} 
\State $r_k^i:=r_k^i/\max_{j=1}^d  (r_k^j)$, $i=1,\dots, d$, $k=1,\dots,n$
\EndIf
\State $\texttt{attempt}:=0$
\While{$\texttt{attempt}\le\texttt{n\_attempts}$}
\State Choose $k\in\{1,\dots,n\}$ uniformly at random
\State Choose $\eta\in [0,\overline\eta]$ uniformly at random
\State $ a^i:=\nu^i \exp\left(\eta r_k^i/\langle\nu, r_k\rangle^{1-\alpha}  \right),\quad w^i:=\frac{a^i}{\sum_{j=1}^d a^j},$
\State $\texttt{attempt}:=\texttt{attempt}+1$
\If{$\frac{1}{n}\sum_{t=1}^n \langle w,r_t\rangle^{\alpha}\ge\frac{1}{n}\sum_{t=1}^n \langle \nu,r_t\rangle^{\alpha}+\texttt{threshold}$}
\State $\nu:=w$, $\texttt{attempt}:=0$
\EndIf
\EndWhile
\Ensure an optimal portfolio $\nu$
 \end{algorithmic}
\end{algorithm}

The algorithm accepts either the original returns $r_k$, or the scaled returns $r_k/r_k^*$. The first case corresponds to the traditional power utility, the second one to the relative power utility. At each point $\nu$ the algorithm tries to make a step according to line 9, corresponding to (\ref{3.2}), where the return $r_k$ and the learning rate are taken randomly by sampling $k$ and $\eta$ from the uniform distributions over $\{1,\dots,n\}$ and $[0,\overline\eta]$ respectively. In fact, this is a step of a stochastic gradient method with random learning rate. That's why we call the algorithm ``doubly stochastic''. Furthermore, the step will be actually performed only if the value of the objective function for the new portfolio $w$ surpasses the current value by a \texttt{threshold}: line 11. The algorithm stops if no such improvement is obtained for some predefined number of attempts: \texttt{n\_attempts}. 

For the logarithmic utility one should put $\alpha=0$, and substitute in line 11 the power function by the logarithm. We do not consider the relative utility in this case. 

The algorithm was applied to NYSE$_1$ and NYSE$_2$ datasets with the following parameters: $\overline\eta=1$, $\texttt{n\_attempts}=10^4$, $\texttt{threshold}=10^{-10}$.
The number of iterations and the results depend on the \texttt{seed} parameter. The average number of attempts to improve the current portfolio for 30 runs of the algorithm was about $283\cdot 10^3$ for NYSE$_1$ and $73\cdot 10^3$ for NYSE$_2$.  In both cases the output portfolio $\nu$ concentrates only on few stocks: 5 for NYSE$_1$ and 3 for NYSE$_2$.
We drop $\nu^i$ with $\nu^i<0.001$ and normalize the results:  
\[ \nu^i:=\frac{\nu^i I_{\{\nu^i\ge 0.001\}}}{\sum_{j=1}^d \nu^j I_{\{\nu^j\ge 0.001\}}}. \]

For the logarithmic utility the results can be compared with those of \cite{Borodin2000,Gyorfi2012}. In Tables \ref{tab:2}, \ref{tab:3} we present minimal and maximal values for each weight, obtained in 30 runs of the GDSEG algorithm. The accumulated wealth $X_n=\prod_{t=1}^n\langle\nu,r_t\rangle$, in fact, does not depend on a particular output $\nu$:
\[ \textrm{NYSE$_1$}: X_{5651}\approx 250.6,\quad \textrm{annual return: } 1.279; \]   
\[ \textrm{NYSE$_2$}: X_{11178}\approx 4100.8,\quad \textrm{annual return: } 1.206.\]
The annual return is computed by the formula $X_n^{252/n}$. 

\begin{table}
\caption{Optimal weights for the logarithmic utility, NYSE$_1$: 30 experiments of the GDSEG algorithm}
\label{tab:2}
\begin{tabular}{ c c c c }
\hline\noalign{\smallskip}
Stock & \makecell{Weight\\ \cite{Borodin2000}} &  \makecell{Weight\\ GDSEG, $[\min,\max]$}  \\
\hline\noalign{\smallskip}
comme & 0.2767  & $[0.2766,0.2770]$  \\ 

espey & 0.1953 &  $[0.1952,0.1956]$ \\ 

iroqu & 0.0927 &  $[0.0925, 0.0929]$ \\  

kinar &  0.2507 &  $[0.2506,0.2508]$ \\

meico &  0.1845 &   $[0.1842,0.1847]$ \\
\end{tabular}
\end{table}

\begin{table}
\caption{Optimal weights for the logarithmic utility, NYSE$_2$: 30 experiments of the GDSEG algorithm}
\label{tab:3}
\begin{tabular}{ c c c   }
\hline\noalign{\smallskip}
Stock & \makecell{Weight \\ \cite{Gyorfi2012}} & \makecell{Weight \\ GDSEG, $[\min,\max]$}   \\
\hline\noalign{\smallskip}
hp  &  0.177 & $[0.1771,0.1776]$  \\ 

morris & 0.747 & $[0.7468,0.7472]$\\ 

schlum & 0.076 & $[0.0753,0.0757]$\\  
\end{tabular}
\end{table}

In general the GDSEG algorithm need not be so stable. For the power utility $u(x)=x^{\alpha}$ 
we implemented the following strategy: take an output $\nu$, corresponding to the largest value of the empirical utility function obtained in 10 experiments. The results for NYSE$_2$ dataset are presented in Table \ref{tab:4}. In the sequel we concentrate only on NYSE$_2$.

\begin{table}[h!] 
\caption{NYSE$_2$: optimal portfolio weights, corresponding to the largest value of the empirical power utility function obtained in 10 experiments of the GDSEG algorithm; the accumulated wealth $X_n$, $n=11178$; the annual returns and the annual volatilities of these portfolios}
\label{tab:4}
\begin{tabular}{ l c | c   c  c  c |  c c c c}
\multicolumn{2}{c|}{} & \multicolumn{4}{c|}{Ordinary utility}            &       \multicolumn{3}{c}{Relative utility}     \\
\hline
$\alpha$     & Stocks                                                 & Weights & $X_n$  & \makecell{Ann\\ ret.} &    \makecell{Ann.\\ volat.} & Weights & $X_n$  & \makecell{Ann.\\ ret.} & \makecell{Ann.\\ volat.}\\
\hline
$0.01$  &     \makecell{hp\\  morris\\  schlum}       &     \makecell{0.1792\\ 0.7518\\ 0.0690} & 4100.4 &   1.206 & 0.234 &\makecell{0.1782\\ 0.7523 \\ 0.0695} & 4100.4 & 1.206 & 0.234\\
\hline
$0.1$   &     \makecell{hp\\  morris\\  schlum}       &     \makecell{0.1762 \\ 0.7766 \\ 0.0473 } & 4091.2 & 1.206 &  0.237  &\makecell{0.1617 \\ 0.7882 \\ 0.0501} & 4085.7 & 1.206 &  0.238\\
\hline 
$0.2$ &  \makecell{hp\\  morris}  &  \makecell{0.1779\\  0.8221}   & 4035.7 & 1.206 & 0.245 & \makecell{0.1476 \\ 0.8524}& 3999.7 & 1.206 & 0.248\\
\hline 
$0.3$ &  \makecell{hp\\  morris}  &  \makecell{0.1589\\  0.8411}   & 4016.1 &  1.206 & 0.247  &\makecell{0.1069\\ 0.8931}  & 3912.5 & 1.205 & 0.253\\
\hline 
$0.5$ &  \makecell{hp\\  morris}  &  \makecell{0.0972\\  0.9028}   & 3885.4 & 1.205 & 0.254 &\makecell{0\\  1} & 3496.7 & 1.202 & 0.270\\
\hline 
$0.75$ &  \makecell{morris}  &  1   &  3496.7 & 1.202 & 0.269  & 1   &  3496.7 & 1.202 & 0.270\\
\end{tabular}
\end{table}

Note that as $\alpha$ is growing, the utility maximizer concentrates more on one stock. This effect is stronger for the relative utility. Such behavior can be qualified as more risky: see the annual volatility of portfolio returns in Table \ref{tab:4}. This quantity is defined as the empirical standard deviation of $(\langle\widehat\nu_n,r_k\rangle)_{k=1}^n$, multiplied by $\sqrt{252}$. For the log-optimal portfolio from Table \ref{tab:3} it equals to 0.233.

Data used in the above calculations can be considered as a realization of some multidimensional stochastic process. From the example considered in Section \ref{sec:5} it is clear that the values of an empirical utility function can be very sensitive to such realizations. To get more insight on the risk and generalization properties of empirically optimal portfolios, let us try to describe the stock prices by the multidimensional Black-Scholes model:
\begin{equation} \label{6.1}
 dS_t^i=S_t^i\mu^i dt +S_t^i\sum_{j=1}^m\sigma^{ij}\,dW_t^j,  \quad i=1,\dots,d, 
\end{equation} 
where $(W^1,\dots,W^m)$ is a standard Wiener process, $\mu$ is the drift vector and $\sigma$ is the volatility matrix. Solving the system of stochastic differential equations (\ref{6.1}), we get 
\[ S_t^i=S_0^i\exp\left(\left(\mu^i-\frac{1}{2}\sum_{j=1}^m(\sigma^{ij})^2\right)t +\sum_{j=1}^m \sigma^{ij}W^j_t\right),\quad i=1,\dots,d. \]
If $t=1$ corresponds to one year, then the daily log-returns should be approximated as follows
\begin{equation} \label{6.2}
 \ln r_k^i=\left(\alpha^i-\frac{1}{2}\sum_{j=1}^m(\sigma^{ij})^2 \right) h+\sum_{j=1}^m \sigma^{ij}(W^j_{kh}-W^j_{(k-1)h}),\quad h=1/252,\quad k=1,\dots,n.
\end{equation} 

We estimated the expectation vector and the covariance matrix 
\[ \left(\alpha^i h -\frac{1}{2}\sum_{j=1}^m(\sigma^{ij})^2 h\right)_{i=1}^d,\qquad \left(\sum_{k=1}^m \sigma^{ik}\sigma^{kj}h\right)_{i,j=1}^d \]
of $(\ln r_k^i)_{i=1}^d$ for NYSE$_2$ dataset, using the \texttt{numpy} module. This allows to generate the artificial data by (\ref{6.2}). For the empirically optimal portfolios from Tables \ref{tab:3}, \ref{tab:4}, as well as for the portfolio with uniform weights: $w=(1/d,\dots,1/d)$, $d=19$, we computed some statistical characteristics of the annual accumulated wealth $X_{252}$, using these data. The results are collected in Table \ref{tab:5}. This table mainly demonstrates the risk properties of empirically optimal portfolios. For example, as $\alpha$ growth, the portfolios become more risky: their expectations and standard deviations increase, but medians decrease. The portfolios, corresponding to the relative power utility are more risky than for the ordinary one, in contrast to the example in Section \ref{sec:5}, but in accordance with Table \ref{tab:4}: see again the annual volatility columns. 

\begin{table}
\caption{Statistical characteristics of the annual accumulated wealth $X_{252}$ for the portfolios from Table \ref{tab:4} for the artificial data (\ref{6.2}) with the parameters, estimated for NYSE$_2$. Averaging was performed over $10^6$ realizations, generated by the Black-Scholes model.}
\label{tab:5}
\begin{tabular}{ c| c | c  | c | c |c}
  Portfolio              & Mean &  Median & \makecell{Std. \\ deviation} & \makecell{5-th\\ percentile} & \makecell{95-th \\ percentile} \\
\hline
uniform   & 1.165 & 1.152 & 0.183 & 0.891 & 1.487 \\
log-optimal  & 1.240 & 1.207 & 0.294 & 0.820  & 1.772 \\
\makecell{$\alpha=0.01$\\ ordinary \\ relative} & \makecell{ \\ 1.240 \\ 1.240} &\makecell{\\ 1.207  \\ 1.207} & \makecell{ \\ 0.295 \\ 0.295} &\makecell{  \\ 0.819  \\  0.819} & \makecell { \\ 1.775 \\ 1.775}\\
\makecell{$\alpha=0.1$\\ ordinary\\ relative} & \makecell{ \\ 1.241 \\ 1.242} & \makecell{ \\ 1.207 \\ 1.207 } & \makecell{ \\ 0.299 \\ 0.300} & \makecell{ \\ 0.815 \\ 0.814} & \makecell{ \\ 1.785\\ 1.787 }\\
\makecell{$\alpha=0.2$\\ ordinary\\ relative} & \makecell{ \\ 1.243 \\ 1.244} & \makecell{ \\ 1.206 \\ 1.206} & \makecell{ \\ 0.310 \\ 0.314} & \makecell{ \\ 0.805\\ 0.801}  & \makecell{ \\ 1.808 \\ 1.815}\\
\makecell{$\alpha=0.3$\\ ordinary\\ relative} &  \makecell{ \\ 1.244\\ 1.245 } & \makecell{ \\ 1.206 \\ 1.205} & \makecell{ \\ 0.312 \\ 0.320} & \makecell{\\ 0.803\\ 0.794} & \makecell{ \\ 1.812 \\ 1.828}\\
\makecell{$\alpha=0.5$\\ ordinary\\ relative} & \makecell{ \\ 1.245 \\1.247} & \makecell{ \\ 1.205 \\  1.202} & \makecell{ \\ 0.322 \\ 0.342} & \makecell{ \\ 0.793\\ 0.771} & \makecell{ \\ 1.831\\ 1.872}
\end{tabular}
\end{table}

The considered dataset is favorable for the investor: the stock prices are growing (on average). Moreover, the performance is evaluated with respect to a concrete model. However, even in this case the investment decisions, based on the historical data, are risky. For example, from Table \ref{tab:5} we see that for the log-optimal portfolio there is 5\% chance to loose more than 18\% of an initial wealth within 1 year. 

Note that  the means are larger than the medians. This is in line with \cite{Gyorfi2012}, where it is explained that typically $X_n$ is less then the $\mathsf E X_n$ for log-optimal portfolios. We see also that the medians give good estimates for the annual returns from Table \ref{tab:4}. 

Finally, we tried to estimate the true utility of the empirically optimal portfolios, constructed for trajectories of the Black-Scholes model. We used the same method as in Section \ref{sec:5}, but with the GDSEG algorithm instead of bisection. Namely, for $\alpha=0.2$ we considered 200 trajectories $(r_1,\dots,r_n)$, $n=11178$ generated by the Black-Scholes model (\ref{6.2}) with parameters, estimated for NYSE$_2$ dataset. For each trajectory the empirically optimal portfolio was computed by the GDSEG algorithm (we picked the best portfolio in 10 experiments).

\begin{figure}[h] 
\centering
\includegraphics[scale=0.95]{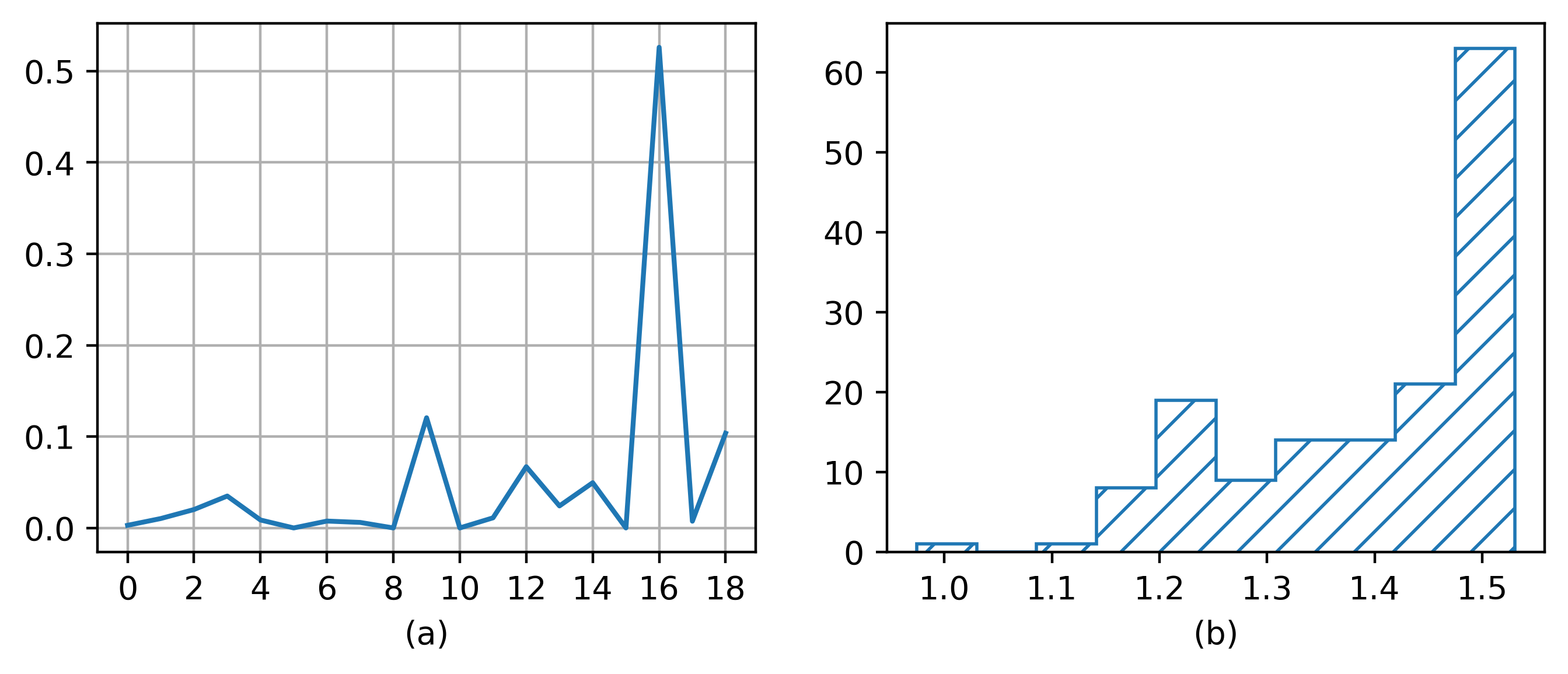}
\caption{Relative power utility with $\alpha=0.2$. (a)  Average weight of each stock in empirically optimal portfolio over 200 realizations of the Black-Scholes model (\ref{6.2}); (b) Histogram of the evaluated true utility for the same 200 optimal portfolios.}
\label{fig:2}
\end{figure}

For a fixed trajectory the optimal portfolio concentrated on a few number of stock (from 1 to 4). For illustration purposes in Fig.\,\ref{fig:2}(a) we show the average weight of each stock over 200 optimal portfolios. As in Table \ref{tab:3}, the largest average weights have the stocks with numbers 9 (hp), 16 (morris), 18 (schlum). The next two positions occupy 12 (jnj) and 14 (merck).

The true utility of each portfolio was evaluated by the empirical mean, computed for a large sample: $n=10^7$. In Fig.\,\ref{fig:2}(b), similar to left panels in Fig.\,\ref{fig:1}, we see a large cluster of very good portfolios. However, the the concentration is far from perfect. Let us mention also that the median ($\approx 1.45$) of the true utility is greater than the mean ($\approx 1.40$).

\section{Conclusion}
In this paper we studied generalization properties of the empirically optimal portfolios for the relative utility maximization problem. We obtained high probability bounds for the estimation error and for the difference between the empirical and true utilities. Similar bounds were obtained for the portfolios, produced by the stochastic exponentiated gradient algorithm. The only assumptions, imposed on the returns is the i.i.d. hypothesis. The obtained bounds depend only the information available to the investor. We also performed some statistical experiments, demonstrating risk and generalization properties of the empirically optimal portfolios. For a multidimensional problem we proposed the greedy doubly stochastic exponentiated gradient (GDSEG) algorithm.

Let us mention some topics for further study. 

\begin{itemize}
\item In Theorems \ref{th:1} -- \ref{th:3} we considered the case of relative utility functions. To obtain similar bounds for ordinary utilities, in general one need to analyze the tails of the return distributions. In addition, the results of \cite{Cortes2019} should be useful for analysis of this problem.
 \item The proposed GDSEG algorithm was enough for our purposes, but it requires large amount of calculations. It may be interesting to study this algorithm and its improvements in more detail. 
\item Using side information is an important method for the construction of successful portfolio strategies. The recent papers \cite{Bertsimas2020,Bazier2020} contain theoretical and practical ideas that can be employed to study this problem in the statistical learning framework.
\end{itemize}

\bibliographystyle{plain}
\bibliography{litPortf}

\end{document}